\documentclass[twocolumn, prl, superscriptaddress, notitlepage]{revtex4-1}
\usepackage{graphicx}
\usepackage{physics}
\usepackage{color}
\usepackage{amsmath}
\usepackage{amsfonts}
\usepackage[normalem]{ulem}
\usepackage{xspace}
\usepackage[dvipsnames]{xcolor}
\usepackage{dcolumn}
\usepackage{textcomp}
\usepackage[breaklinks,colorlinks,linkcolor=blue,citecolor=blue,urlcolor=blue]{hyperref}

\begin{document}
\title{Building Krylov complexity from circuit complexity}

 \author{Chenwei Lv}
\thanks{They contribute equally to this work.}
 \affiliation{Department of Physics and Astronomy, Purdue University, West Lafayette, IN, 47907, USA}

 \author{Ren Zhang}
\thanks{They contribute equally to this work.}
\affiliation{School of Physics, Xi'an Jiaotong University, Xi'an, Shaanxi 710049}

\author{Qi Zhou}
\email{zhou753@purdue.edu}
\affiliation{Department of Physics and Astronomy, Purdue University, West Lafayette, IN, 47907, USA}
\affiliation{Purdue Quantum Science and Engineering Institute, Purdue University, West Lafayette, IN 47907, USA}
\date{\today}

\begin{abstract}
Krylov complexity has emerged as a new probe of operator growth in a wide range of non-equilibrium quantum dynamics. However, a fundamental issue remains in such studies: the definition of the distance between basis states in Krylov space is ambiguous. 
Here, we show that Krylov complexity can be rigorously established from circuit complexity when dynamical symmetries exist. 
Whereas circuit complexity characterizes the geodesic distance in a multi-dimensional operator space, Krylov complexity measures the height of the final operator in a particular direction. 
The geometric representation of circuit complexity thus unambiguously designates the distance between basis states in Krylov space. 
This geometric approach also applies to time-dependent Liouvillian superoperators, where a single Krylov complexity is no longer sufficient. 
Multiple Krylov complexity may be exploited jointly to fully describe operator dynamics. 
\end{abstract}
\maketitle 

In the past few years, significant progress has been made in the study of non-equilibrium quantum phenomena, ranging from eigenstate thermalization to  quantum information scrambling~\cite{Nonequilibrium2011, Altman2015, Heyl2013, Heyl2018, Maldacena1999, DAlessio2016, Maldacena2016, Hartnoll2018}. 
A central task of these studies is to explore how local quantum information spreads to the vastly large degree of freedom in a many-body system~\cite{Sekino2008, Lashkari2013, Shenker2014, Maldacena2016}. 
Some commonly used theoretical tools include the out-of-time-order correlator and Loschmidt echoes, among the many others~\cite{Swingle2018, Roberts2018, Magn2018, Qi2019, Zurek2020, ChaosComplexity2020, Magn2020}. 
Krylov (K-) complexity has recently been introduced as a new probe of quantum information spreading in non-equilibrium dynamics~\cite{Altman2019}. 
In the Heisenberg picture, a local operator may evolve into a non-local one in quantum dynamics. K-complexity traces such operator size growth and thus provides physicists with deep insights to many fundamental questions in non-equilibrium dynamics~\cite{Altman2019, Dymarsky2021, Rabinovici2021, Bhattacharjee2022, Caputa2022, Rabinovici2022, Zhai2022, Vijay2022}.    

The study of K-complexity is based on the Lanczos algorithm~\cite{Lanczos1950, Altman2019}. 
In the Heisenberg picture, an operator is denoted by $\hat{\mathcal{O}}(t)=\sum\nolimits_{m,n}\mathcal{O}_{mn}\ket{m}\bra{n}$, where $t$ is the time, $\ket{m}$ is a set of orthogonal eigenstates and $\mathcal{O}_{mn}$ are the matrix elements. $\hat{\mathcal{O}}(t)$ may be regarded as a state in the operator space and thus can be denoted as $|\mathcal{O}(t))\equiv\sum\nolimits_{m,n}\mathcal{O}_{mn}\ket{m}\otimes\ket{n}$. 
Using the Krylov basis $|\mathcal{O}_n)$, an ordered set of mutual orthogonal operators,   
\begin{equation}
|\mathcal{O}(t))=\sum_n \phi_n(t)|\mathcal{O}_n),\label{eq1}
\end{equation} 
where $|\mathcal{O}_{0})=|\mathcal{O}(t=0))$. $|\mathcal{O}_{n\neq0})$ is generated by commutators of $\hat{\mathcal{O}}_0$ and $\hat{H}$ recursively using the Lanczos algorithm. 
For instance, $|\mathcal{O}_1)=b_1^{-1}|[H,\mathcal{O}_0])$, where the normalization constant $b_1$ is introduced such that $(\mathcal{O}_n|\mathcal{O}_n)=(\mathcal{O}_0|\mathcal{O}_0)$. 
The inner product in operator space is defined as $(A|B)=\Tr(A^\dag B)$.
The other Krylov basis states are defined recursively by $|\mathcal{O}_n)=b_n^{-1}(|[H,\mathcal{O}_{n-1}])-b_{n-1}|\mathcal{O}_{n-2}))$. 
K-complexity is defined as 
\begin{equation}
C_K(t)=\sum_n n|\phi_n(t)|^2\label{KC},
\end{equation}
which can be viewed as the expectation value of the Krylov operator 
\begin{equation}
\hat{\cal K}_\mathcal{O}=\sum_{n} n |\mathcal{O}_n)(\mathcal{O}_n|.\label{eq:KrylovOperator}
\end{equation}
Intuitively, Eq.(\ref{KC}) describes the mean width of a wavepacket in the Krylov space and thus quantitatively measures how the size of the operator increases as time goes by. 
Whereas it was hypothesized that $C_K$ has the fastest growth in a chaotic system~\cite{Altman2019}, recent studies have shown that similar behaviors of $C_K$ may arise in non-chaotic systems~\cite{Bhattacharjee2022, Caputa2022}.   

Despite the exciting development in the study of K-complexity, a fundamental question remains. 
The Lanczos algorithm only provides an ordered Krylov basis but does not supply a distance among them.  
To obtain the operator size described by Eq.(\ref{KC}), $|\mathcal{O}_n)$ are assumed to be equally spaced. 
As such, any quantitative results of K-complexity are built upon the choice that the distance between $|\mathcal{O}_n)$ and $|\mathcal{O}_0)$ has been chosen as $n$. 
In practice, substituting $n$ by any other function $h(n)$ shall change all results of K-complexity. 
At a more fundamental level, without prior knowledge of the geometry of the operator space, any choice of the distance between $|O_n)$ could be regarded as ambiguous. 
It is thus desirable to define K-complexity rigorously by specifying the geometry of the Krylov space where $|\mathcal{O}_n)$ live.  

Here, we show K-complexity can be systematically established from circuit (C-) complexity, a concept that originated from quantum computation and is now being widely used in many other areas including high energy physics and condensed matter physics~\cite{Nielsen2006, Jefferson2017, Chapman2018, Guo2018, Chapman2019, Susskind2019}. 
In quantum computation, C-complexity describes the smallest number of gates to reach a target state from a reference state. 
It can be visualized using Nielsen's geometric approach, which provides a metric for operators such that C-complexity is given by the geodesic in the space of circuits~\cite{Nielsen2006}. 
Compared to other definitions of the metric in operator space, such as the trace distance~\cite{Wilde2013}, a unique advantage of circuit complexity is that the distance between operators has a clear physical interpretation. 
A longer(shorter) path corresponds to more (less) gates required to evolve from one operator to the other.  
Applying circuit complexity to non-equilibrium quantum dynamics, it can be understood as the shortest time to reach a desired state and thus allows experimentalists to optimize quantum controls~\cite{Zhou2022}.   

\begin{figure}
    \centering
    \includegraphics[width=0.45\textwidth]{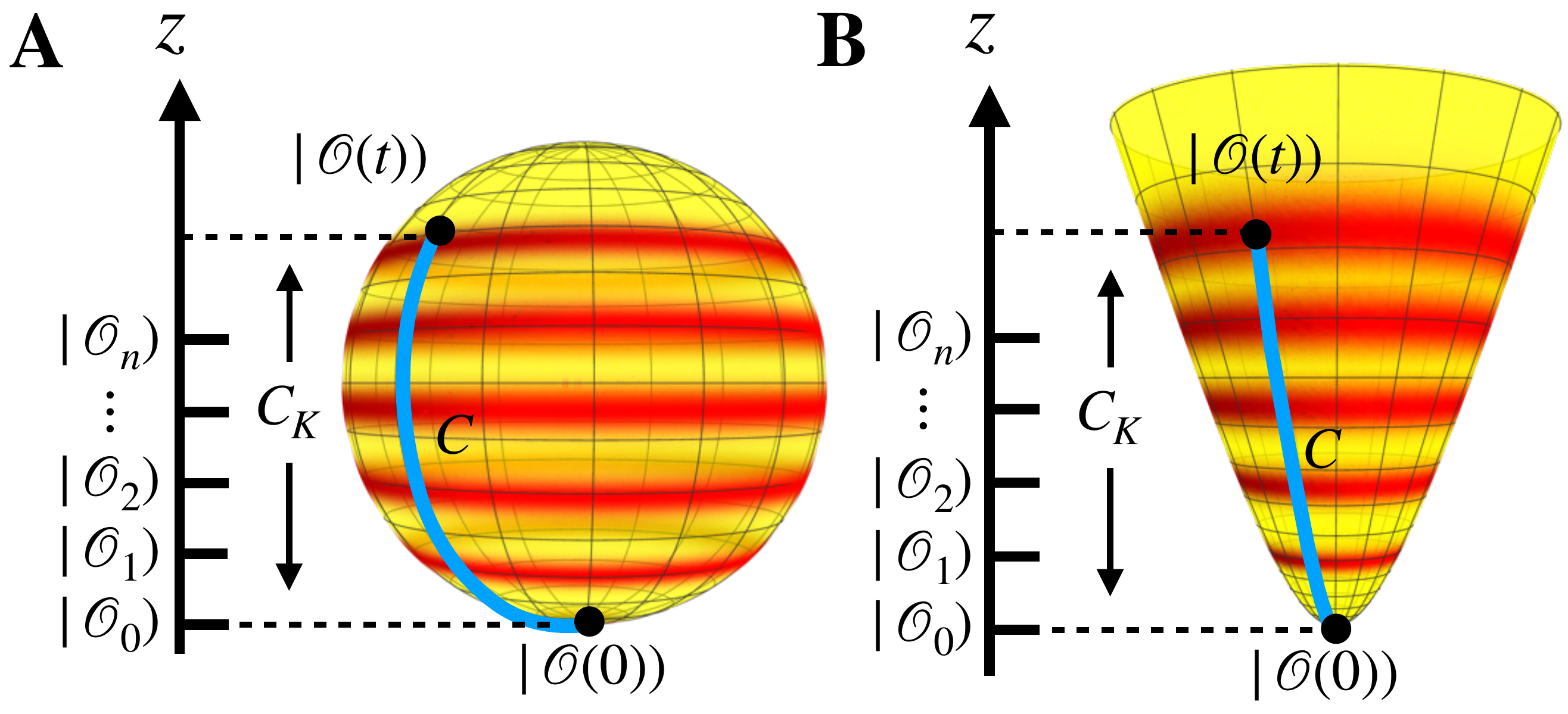}
    \caption{
    A schematic of the Krylov basis as strips on a 2-sphere (A) and a hyperboloid (B) for SU(2) and SU(1,1), respectively. 
    The curves on the surfaces denote the geodesics and the geodesic length is state C-complexity $C$. 
    The height of the final operator is K-complexity $C_K$. }  
    \label{fig:fig1}
\end{figure}

We have found that K-complexity measures the height of the time-dependent operator along a particular direction in the operator space. 
For instance, in systems with SU(2) symmetry, the metric of all operators involved in quantum dynamics forms a sphere. 
As shown in Fig.~\ref{fig:fig1} (A), the initial operator $|\mathcal{O}(0))$ is placed at the south pole.  
The length of a trajectory starting from the south pole along a big circle provides us with the circuit complexity. 
Projecting this trajectory to the $z$-axis, the K-complexity is the height of the final operator measured from the south pole. 
Our results also show that the distance between Krylov basis states is rigorously determined by the circuit complexity. 
Here, the Krylov basis states correspond to strips along latitudes on the sphere. 
Such a distance turns out to be variable once it becomes easier or more difficult to change the operators in some directions in the operator space. 

We note that the geometric interpretation of K-complexity has recently been studied in an elegant work by Caputa et al~\cite{Caputa2022}. 
It was argued that the K-complexity corresponds to the area of a certain region, using the Fubini-Study metric to define the metric of the operator space~\cite{QuantumMetric2017, QuantumMetric2019, QuantumMetric2020}.
Compared to such an inner product of operators, the physical meaning of C-complexity used here is more clear in quantum dynamics. 
We find that it is more appropriate to interpret K-complexity as a length rather than an area. 
Furthermore, our approach applies to time-dependent Liouvillian
superoperators, a largely unexplored problem in the study of K-complexity. 
As we will show, it requires multiple K-complexity to fully describe the operator growth in the most generic case where Liouvillian
superoperators are time-dependent.

In the operator space, the Heisenberg equation of motion is recast into a similar form as the Schr\"odinger equation,
\begin{equation}
    -i\partial_t|\mathcal{O}(t))=\hat{\mathcal{L}}|\mathcal{O}(t)).\label{mal}
\end{equation}
where the Liouvillian superoperator becomes $\hat{\mathcal{L}}=\hat{H}\otimes\hat{\mathbb{I}}-\hat{\mathbb{I}}\otimes \hat{H}^T$.
When a dynamical symmetry exists, this equation has simple analytical solutions. 
For instance, when $\hat{\mathcal{L}}$ is written as $\hat{\mathcal{L}}=B\hat{\mathcal{S}}_x$,
\begin{equation}   
    \hat{\mathcal{S}}_x=\sum_{n=0}^{2\ell-1} b_n|\mathcal{O}_{n+1})(\mathcal{O}_{n}|+{\rm h.c.}\label{mL}
\end{equation}
where $b_n= \sqrt{(n+1)(2\ell-n)}$, $2\ell$ is an integer denoting the total number of Krylov basis states, and $B$ is a constant. 
One can define $\hat{\mathcal{S}}^+=\sum\nolimits_n b_n|\mathcal{O}_{n+1})(\mathcal{O}_{n}|$ and $\hat{\mathcal{S}}^-=\sum\nolimits_n b_n|\mathcal{O}_{n})(\mathcal{O}_{n+1}|$. $\hat{\mathcal{K}}_\mathcal{O}$ and $\hat{\mathcal{S}}^{\pm}$ provide three generators of SU(2), as recognized by Caputa et al~\cite{Caputa2022}. 
\begin{equation}
   [\hat{\mathcal{K}}_\mathcal{O},\hat{\mathcal{S}}^\pm]=\pm \hat{\mathcal{S}}^\pm,\quad [\hat{\mathcal{S}}^+,\hat{\mathcal{S}}^-]=2(\hat{\mathcal{K}}_\mathcal{O}-\ell). 
\end{equation}
This dynamical symmetry can be easily seen by noting that Eq.(\ref{mL}) corresponds to the spin operator $\hat{\mathcal{S}}_x$ of a total spin-$\ell$, $\hat{\mathcal{S}}_y=(\hat{\mathcal{S}}^+-\hat{\mathcal{S}}^-)/(2 i)$ and $\hat{\mathcal{S}}_z=\hat{\mathcal{K}}_\mathcal{O}-\ell$, and the constant $B$ corresponds to the strength of the magnetic field acting on the spin. Eq.(\ref{mal}) thus can be viewed as the Schr\"odinger equation for a spin-$\ell$ subject to a constant magnetic field in the $x$-direction. Alternatively,  $\phi(t)$ in Eq.(\ref{eq1}) can be regarded as the time-dependent wavefunction in a lattice model where the tunneling amplitude is denoted by $b_n$.

For Liouvillians with dynamical symmetry, the propagator 
$\hat{\mathcal{U}}=\mathcal{T}e^{i\int\hat{\mathcal{L}}dt}$, where $\mathcal{T}$ is the time-ordering operator, can be parameterized using the generators of the symmetry group in the same manner as the evolution of a quantum state. For SU(2), 
\begin{equation}
    \hat{\mathcal{U}}=e^{i\psi \hat{\mathcal{S}}_z}e^{i\theta \hat{\mathcal{S}}_x}e^{i\varphi \hat{\mathcal{S}}_z}.\label{pro}
\end{equation}
It evolves the initial operator $|\mathcal{O}(0))$ to a generalized coherent state $|\mathcal{O}(t))=\hat{\mathcal{U}}|\mathcal{O}(0))=|\mathcal{C}_{\theta,\varphi,\psi})$~\cite{Gilmore1990}. 
Eq.(\ref{pro}) allows us to define circuit complexity and its underlying geometry. 

In the most generic case, the Liouvillian superoperator in Eq.~\ref{mal} is given by $\hat{\mathcal{L}}=B_x\hat{\mathcal{S}}_x+B_y
\hat{\mathcal{S}}_y+B_z\hat{\mathcal{S}}_z $. 
The operator dynamics thus can be viewed as the evolution of a spin-$\ell$ subject to a magnetic field $\vec{B}(t)=(B_x, B_y, B_z)$. 
The circuit complexity denotes the shortest time to reach a target state under the constraint that the strength of the magnetic field is fixed, i.e., $|\vec{B}|=\sqrt{B_x^2+B_{y}^2+B_z^2}\equiv B$~\cite{Zhou2022}. 
This amounts to fixing the metric of the operator space ~\cite{Bengtsson2017, Susskind2019},
\begin{equation}
    ds^2=\ell^2 dt^2(B_x^2+B_y^2+B_z^2).\label{costI}
\end{equation}
Noting that $d\hat{\mathcal{U}}\hat{\mathcal{U}}^{-1}=i \hat{\mathcal{L}} dt$, equations (\ref{pro})
shows that $d\theta$, $d\psi$ and $d\varphi$ are determined by $\vec{B}$. Using $\text{Tr}(\mathcal{\hat{S}}_i\mathcal{\hat{S}}_j^\dagger)=\delta_{i,j}$, $B_{i=x,y,z}$ can be expressed in terms of $\psi$, $\theta$, $\varphi$, $d\psi$, $d\theta$ and $d\varphi$.
In the content of circuit complexity, it amounts to defining a cost function $F^2=\ell^2{\rm Tr}(d\hat{\mathcal{U}}\hat{\mathcal{U}}^{-1}(d\hat{\mathcal{U}}\hat{\mathcal{U}}^{-1})^\dag )$~\cite{Susskind2019}.
Substituting such expressions to Eq.(\ref{costI}), we obtain the following metric
\begin{equation}
ds^2= \ell^2(d\theta^2+d\psi^2+d\varphi^2+2\cos(\theta)d\varphi d\psi). \label{3Sphere}
\end{equation}
This is the metric of a 3-sphere. 

We note that Eq.(\ref{eq1}) can be understood as an expansion of the coherent state using  $|\mathcal{O}_n)$, which plays the roles of eigenstates of $\hat{\mathcal{S}}_z$ in the usual spin problem.
To be explicit, 
\begin{equation}
    \phi_n(t)=e^{-i\ell (\varphi+\psi)}\cos^{2\ell}(\theta/2)\sqrt{\frac{(2\ell)!}{n!(2\ell-n)!}}\mu^n,
\end{equation}
where $\mu=i\tan(\theta/2)e^{i\psi}$.
As such, {$\varphi$} controls the global phase, {$\psi$} denotes the relative phase between $\phi_n$.
And $\theta$ determines the amplitude distribution of $|\phi_{n}|^2$. 
Very often, two operators with the same global phase can be identified as the same one. 
As such, the state C-complexity is defined as the minimum of C-complexity to the final state with different $\varphi$ but the same $\psi$ and $\theta$. 
Since Eq.~\ref{3Sphere} is a quadratic form of $d\varphi$, extremizing Eq.~\ref{3Sphere} by setting $d\varphi=-\cos(\theta)d\psi$ reduces a 3-sphere to a 2-sphere 
\begin{equation}
    ds^2= \ell^2(d\theta^2+\sin^2(\theta)d\psi^2). \label{2Sphere}
\end{equation}
Each point on the 2-sphere is equivalent to a coherent state $|\mathcal{C'}_{\theta,\psi})\equiv|\mathcal{C}_{\theta,\psi,0})$. 
Since the geodesics of a 2-sphere are big circles, the state C-complexity could be visualized as the length of the arc along a big circle connecting the initial and the final operators, or equivalently, the polar angle $\theta$, as shown in Fig.~\ref{fig:fig1}. 

Since $|\mathcal{C'}_{\theta,\psi})$ form an overcomplete basis, we could expand $|\mathcal{O}_n)$ using $|\mathcal{C'}_{\theta,\psi})$,
\begin{equation}
    |\mathcal{O}_n)=\frac{2\ell+1}{4\pi}\int\sin(\theta) d\theta d\psi f_n(\theta,\psi)|\mathcal{C'}_{\theta,\psi}),
\end{equation}
in the same manner as expanding the eigenstates of $\hat{\mathcal{S}}_z$ using spin coherent states, where $|f_n(\theta,\psi)|^2=|\phi^*_n|^2_{\varphi=0}$ is centered around $\theta_n=\arccos(1-n/\ell)$. 
This is precisely the Husimi Q-representation of $|\mathcal{O}_n)(\mathcal{O}_n|$ using the coherent state of operators~\cite{Gilmore1990}.
On the 2-sphere, $|\mathcal{O}_0)$ and $|\mathcal{O}_{2\ell})$ are placed at the south and the north poles, respectively.   
Each other $|\mathcal{O}_{n\neq 0})$ corresponds to a strip along a latitude with a finite width. 
The projection of the strip to the $z$-axis is equally spaced and the height of the strip measured from the south pole linearly increases with $n$. 
In other words, each $|\mathcal{O}_n)$ can be assigned a unique coordinate on the $z$-axis, $n$. 
This is directly a consequence of the metric tensor in Eq.(\ref{2Sphere}). We conclude that once the metric of the operator space is fixed in the study of C-complexity, the distance between basis states in the Krylov space is uniquely determined. 

Since the weight of $|\mathcal{O}(t))$ in $|\mathcal{O}_n)$ is given by $|\phi_n(t)|^2$, K-complexity defined in Eq.(\ref{KC}) could be viewed as the height of $|\mathcal{O}(t))$ measured from the south pole, in the same manner as the expectation value of $\hat{\mathcal{S}}_z$ of a spin. 
This is the physical meaning of K-complexity in the geometric representation of C-complexity. For the dynamics generated by $\hat{\mathcal{L}}=B\hat{\mathcal{S}}_x$ in Eq.(\ref{mL}), the evolution of the operator is equivalent to a spin processing about the $x$-axis. 
The state C-complexity, which equals the polar angle $\theta$, grows linearly as a function of time, $C=\ell\theta(t)=\ell Bt$. 
K-complexity, the height of the final operator is equivalent to the average value of $\hat{\mathcal{S}}_z$ of a spin, $C_K=\ell-\ell\cos(B t)$. 

The same discussions apply to SU(1,1), where $b_n$ in Eq.~\ref{mL} is replaced by $\sqrt{(n+1)(2k+n)}$, and $k$ is the Bargmann index, a counterpart of $\ell$ in SU(2). 
The state C-complexity could be visualized using a hyperbolic surface embedded in Minkowski space. 
As shown in Fig.~\ref{fig:fig1}(B), each point on the hyperbolic surface is an SU(1,1) coherent state in the operator space,
\begin{equation}
    |C_{\theta,\psi}')=\frac{e^{ik\psi}}{\cosh^{2k}(\theta/2)}\sum_n \sqrt{\frac{\Gamma(2k+n)}{\Gamma(2k)n!}}\tilde{\mu}^n|\mathcal{O}_n)
\end{equation}
where $\tilde\mu=i\tanh(\theta/2)e^{i\psi}$.
$C$ is the length of the arc connecting the initial and final operators on the hyperbolic surface and grows linearly as a function of time, $C=kBt$. The Krylov basis states again correspond to strips on this surface. 
On the $z$-axis, $|\mathcal{O}_n)$ are equally spaced, and $C_K$ is the height of the final operator measured from the initial operator placed at the bottom of the hyperbolic surface. 
When C-complexity grows linearly, the height of the final operator grows exponentially. 
This provides a geometric interpretation of the exponentially growing K-complexity,
\begin{equation}
   C_K(t)= k\cosh(B t)-k.
\end{equation}

%---------------------------------------------------------------
%---------------------------------------------------------------

The metric of operator space, for instance, the 2-sphere in Eq.(\ref{2Sphere}), depends on $ds$ defined in Eq.(\ref{costI}). 
In the study of C-complexity, it is known that the metric of the operator space could  change if a different $ds$ is used. 
Physically, this originates from that it may be easier (or more difficult) to implement certain gates than others~\cite{Susskind2019}. 
For instance, in SU(2), one may choose
\begin{equation}
 ds^2=\ell^2 dt^2(B_x^2+B_y^2+\lambda^2B_z^2).
\end{equation}
$\lambda>1$ ($\lambda<1$)  means that it is more difficult (easier) to implement the gate in the $z$-direction $\hat{\mathcal{L}}_z$. An alternative understanding is that one seeks the least time to access the target state under a different constraint. 
The new choice means that $\sqrt{B_x^2+B_y^2+\lambda^2B_z^2}$ other than $|B|$ is fixed. The metric of the operator space becomes 
\begin{equation}
  \begin{split}
      ds^2=&\ell^2(d\theta^2+d\psi^2+d\varphi^2+2\cos(\theta)d\varphi d\psi)\\&+\ell^2(\lambda^2-1)(\cos(\theta)d\varphi+d\psi)^2.
  \end{split}
\end{equation}
When $\lambda=1$, it reduces to the previous result of a 3-sphere in Eq.(\ref{3Sphere}). 

\begin{figure}
    \centering
    \includegraphics[width=0.495\textwidth]{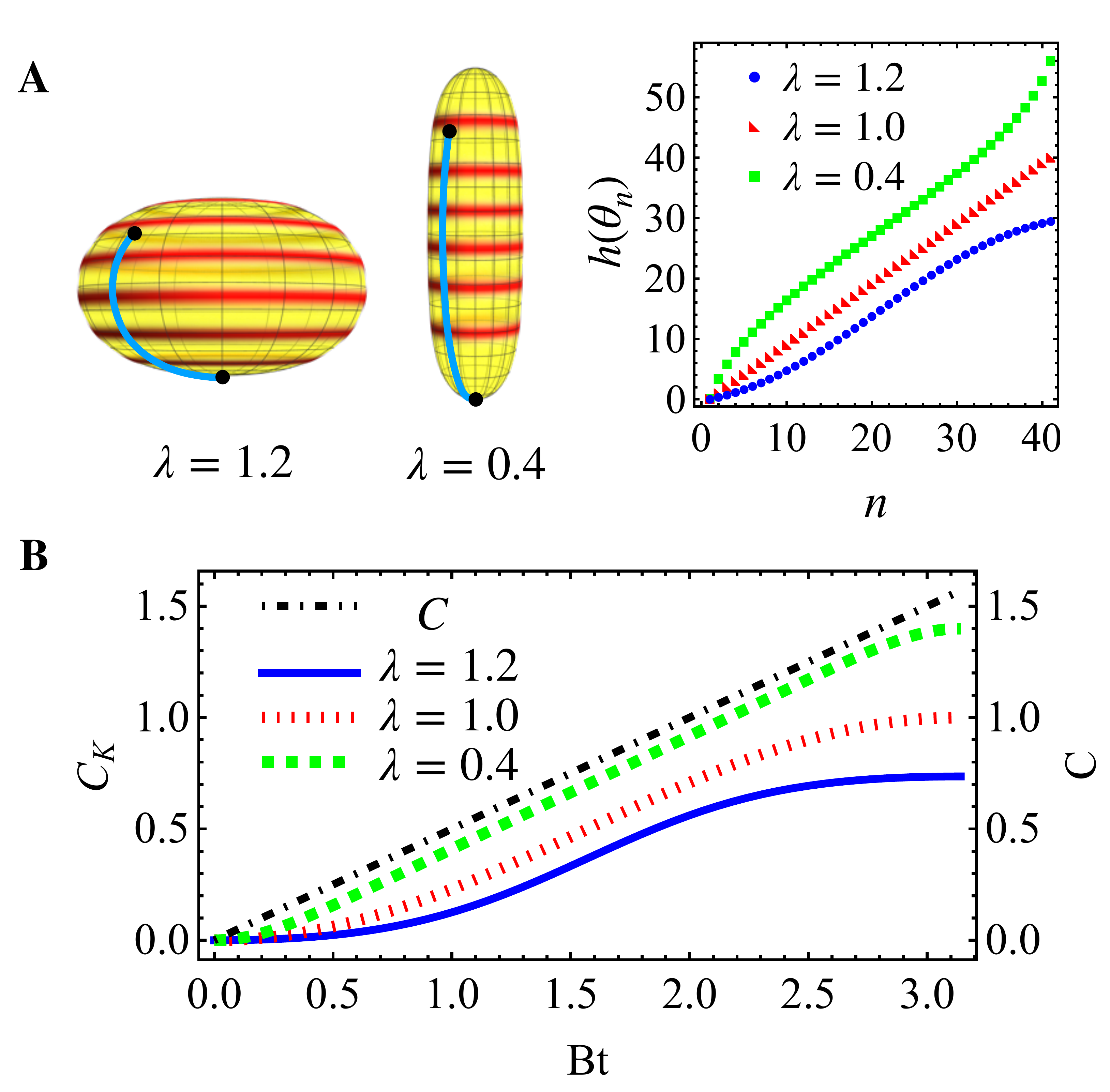}
    \caption{
    (A) Squashed ($\lambda>1$) or stretched ($\lambda<1$) spheres for SU(2). 
    The Krylov basis has unequally spaced heights $h(\theta_n)$ when $\lambda\neq 1$. 
    (B) K-complexity (in unit of $2\ell$) as a function of time $t$ depends on how the sphere is deformed.
    C-complexity always grows linearly. }
    \label{fig:fig2}
\end{figure}

The state C-complexity is obtained by minimizing the above metric, 
\begin{equation}
  d\varphi=-\frac{\lambda^2\cos(\theta)}{1+(\lambda^2-1)\cos^2(\theta)}d\psi.
\end{equation}
We obtain 
\begin{equation}
  ds^2= \ell^2\bigg(d\theta^2+\frac{\lambda^2\sin^2(\theta)}{1+(\lambda^2-1)\cos^2(\theta)}d\psi^2\bigg).\label{bergerSphere}
\end{equation}
When $\lambda=1$, Eq.(\ref{bergerSphere}) reduces to the previous result of a 2-sphere in Eq.(\ref{2Sphere}). 
When $\lambda\neq 1$, Eq.(\ref{bergerSphere}) describes a deformed sphere, as shown in Fig.~\ref{fig:fig2}A. 
The height of a point on the sphere with a polar angle $\theta$ measured from the south pole now becomes
\begin{equation}
  h(\theta)=\ell\int_{0}^{\theta} \sqrt{1-\frac{\cos^2(\theta')}{(\lambda^{-2}\sin^2(\theta')+\cos^2(\theta'))^3}}d\theta'
\end{equation}
When $\lambda\neq 1$, the Krylov basis states $|\mathcal{O}_n)$ are no longer equally spaced. 
$C_K(t)$ in Eq.~(\ref{KC}) needs to be modified such that it gives rise to the height of the final operator $|\mathcal{O}(t))$,
\begin{equation}
    C_K(t)=\sum_n h(\theta_n) |\phi_n(t)|^2.
\end{equation}
As a result, the time dependence of $C_K$ changes, as shown in Fig.~\ref{fig:fig2}B. 

%---------------------------------------------------------------
%---------------------------------------------------------------

\begin{figure}
    \centering
    \includegraphics[width=0.49
    \textwidth]{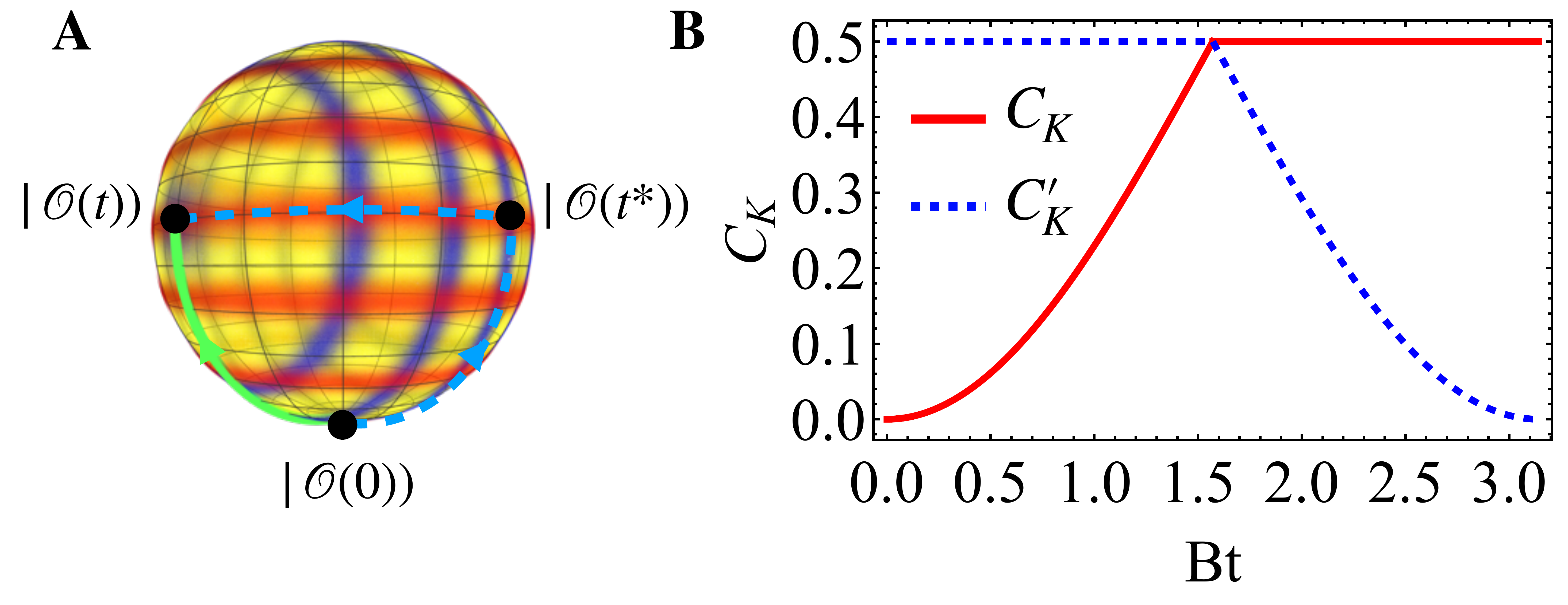}
    \caption{
    (A) The operator dynamics (blue dashed arrows) when the Liouvillian superoperator is time-dependent, changing from 
     $\hat{\mathcal{L}}=B\hat{\mathcal{S}}_x$ to $\hat{\mathcal{L}}'=B\hat{\mathcal{S}}_z$ at $Bt^*=\pi/2$.        
    The red and blue strips on the sphere denotes $|\mathcal{O}_n)$ and $|\mathcal{O}'_n)$, respectively. 
    The green curve with an arrow represents the shortest path connecting the initial and final operators.
    (B) The K-complexity (in the unit of $2\ell$) as functions of time of the quench dynamics.
    }
    \label{fig:fig3}
\end{figure}

Our geometric approach also applies to time-dependent Liouvillian superoperators. 
The simplest case is a quantum quench, where the dynamics is determined by $\hat{\mathcal{L}}$ and $\hat{\mathcal{L}}'$ before and after $t^*$, respectively. 
For instance, in the case of SU(2),  $\hat{\mathcal{L}}=B\hat{\mathcal{S}}_x$ ($\hat{\mathcal{L}}'=B\hat{\mathcal{S}}_z$) when  $t<t^*$ ($t>t^*$). 
The trajectory on the 2-sphere is no longer along the longitude but the latitude when $t>t^*$. 
For simplicity, we consider only undeformed spheres. 
The generalization to $\lambda\neq 1$ is straightforward. 
Though the operator $\mathcal{O}(t)$ is still evolving, $C_K$ remains a constant and thus cannot capture the quantum dynamics when $t>t^*$. 
This can be simply seen from the definition of $C_K$ in Eq.(\ref{KC}), in which only the amplitude of $\phi_n$ is included. 
Nevertheless, the relative phase between $\phi_n$ must be included, since the full dynamics happens on a 2-sphere and the polar angle $\psi$ of this 2-sphere is precisely the relative phase between $\phi_n$. 
State C-complexity $C$ that fully exploits the metric of a 2-sphere can naturally capture the dynamics under an arbitrary time-dependent Liouvillian superoperator. 
For instance, the shortest path connecting the initial and final operators in this quench dynamics is along another longitude, as shown in Fig.~\ref{fig:fig3}A. 
Apparently, this path can be accessed using another time-independent Liouvillian superoperator $\hat{\mathcal{L}}''=B\hat{\mathcal{S}}_y$.

Since a single K-complexity is no longer sufficient when the Liouvillian superoperator is time-dependent, multiple K-complexity may be used. 
In SU(2), we can define another set of Krylov basis states,
\begin{equation}
    |\mathcal{O}'_n)=e^{-i\pi \hat{\mathcal{S}}_y/2} |\mathcal{O}_n),
\end{equation}
which play the same role as eigenstates of $\hat{\mathcal{S}}_x$ in a spin problem. 
On the 2-sphere, $|\mathcal{O}'_n)$ corresponds to another set of strips, as shown in Fig.~\ref{fig:fig3}A. 
Correspondingly, $|\mathcal{O}(t))$ may be expanded using  $|\mathcal{O}'_n)$ as $|\mathcal{O}(t))=\sum_n\phi'_n(t)|\mathcal{O}'_n)$, and another K-complexity is defined $C_K'=\sum_n n|\phi'_n|^2$.
As shown in Fig.\ref{fig:fig3}B, when $t<t^*$, $C_K$ increases with increasing $t$ while $C_K'$ remains unchanged. 
After $t>t^*$, whereas $C_K$ stops growing, $C_K'$ begins to change. 
As such, at any time $t$, $C_K$ and $C_K'$ could jointly determine $|\mathcal{O}(t))$ on the 2-sphere, in the same manner as uniquely determining a spin on the Bloch sphere using the expectation values of both $\hat{S}_z$ and $\hat{S}_x$ in spin tomography~\cite{tomography2}. 
Similar results can be straightforwardly obtained for $SU(1,1)$.

%---------------------------------------------------------------
%---------------------------------------------------------------

In conclusion, we have shown that K-complexity can be rigorously established from C-complexity in the presence of dynamical symmetry. 
Our method provides a clear geometrical picture and physical meaning of K-complexity. 
Our results can be generalized to other symmetry groups, which are expected to bring even richer physics bridging geometry and quantum information spreading in non-equilibrium dynamics.  

Q.Z. acknowledges useful discussions with Pawel Caputa.
This work is supported by DOE DE-SC0019202, and the U.S. Department of Energy, Office of Science through the Quantum Science Center (QSC), a National Quantum Information Science Research Center. 
RZ is supported by NSFC (Grant No.12074307) and National Key R$\&$D Program of China (Grant No. 2018YFA0307601).

% \bibliographystyle{apstest}
% \bibliography{Krylov.bib}


\end{document}